\documentclass[12pt,preprint]{aastex}

\shorttitle{LSB edge-on galaxies}
\shortauthors{Bizyaev and Kajsin}

\begin{document}

\title{Surface photometry of LSB edge-on galaxies}

\author{Dmitry Bizyaev\altaffilmark{1,2}} \and
\author{Serafim Kajsin\altaffilmark{3}}

\altaffiltext{1}{Physics Department, University of Texas at El Paso, TX
79968}
\altaffiltext{2}{Sternberg Astronomical Institute, Moscow, 119899, Russia,
dmbiz@sai.msu.ru}
\altaffiltext{3}{Special Astrophysical Observatory of RAS, pos.
Nizhnij Arkhyz, 357147,  Karachaevo-Cherkessia, Russia, skaj@rebus.sao.ru}

\begin{abstract}

We present results of surface photometry for eleven edge-on galaxies
observed with the 6-m telescope at the Special Astrophysical Observatory of
the Russian Academy of Sciences. The photometric scale length, the scale
height, and the central surface brightness of the stellar disks for the
galaxies were found using photometric cuts made parallel to the major and
the minor axes for each galaxy. We show that four galaxies in our sample
that were visually classified as objects of lowest surface brightness in the
Revised Flat Galaxies Catalog have LSB (low surface brightness) disks.
Stellar disks of our LSB galaxies are thinner than HSB (high surface
brightness) ones. There is a good correlation between the central surface
brightness of the stellar disk and its ratio of vertical to radial scales.
The ratio of the disk photometric scales enables us to estimate the mass of
the spherical galactic subsystem using results from numerical modeling.
Combining our results with published rotation curves we determined the mass
of dark halos for the galaxies in our sample. The LSB galaxies tend to
harbor relatively more massive spherical subsystems than those of HSB's.
Indeed, we found no systematic difference between our LSB and HSB objects in
dark halo masses considering it separately from the bulge. At the same time,
the inferred mass/luminosity ratio for the LSB disks appears to be
systematically higher, when compared to the ratio for HSB ones.

\end{abstract}

\section{Introduction}

One of the main features of the low surface brightness galaxies (LSB) which
make them different from the "usual" high surface brightness (HSB) galaxies
is that they are considered to be dark matter dominated objects.

We conducted the study of a sample of several edge-on LSB and HSB galaxies
in order to compare their photometric parameters (including the stellar disk
thickness) and the relative mass of their dark halos.

\section{Sample of galaxies and observations}

Our sample consists of several objects taken from the Revised Catalog of
Flat Galaxies ([1], RFGC hereafter). All those galaxies included to the
catalog are highly inclined objects. We selected four objects of faintest
surface brightness class as candidates to the LSB galaxies (according to
[2]), and seven objects with intermediate and high surface brightness class
as "reference" HSB objects.

The surface photometry observations were made with the 6-m telescope at the
Special Astrophysical Observatory (Russia). The large aperture of the
telescope provided us with high angular resolution in images (0".2/pix)
together with a good sensitivity to the very faint regions of the galaxies.
The chosen galaxies and the set of calibrating frames were taked during one
observing run in December 2000 using R Johnson-Cousins photometric band.

The combined and calibrated R images were used to obtain the radial scale
length of a galactic disk ($h$), the disk vertical scale heights ($z_0$,
assuming $sech^2$ law), the "face-on" central surface brightness ($\mu_0$) of 
the disk (see [3] for the details), and the bulge to disk luminosity ratio
$Lb/Ld$. The final value of $\mu_0$ was corrected for the extinction in our
Galaxy (according to the LEDA). The fitting profiles have been convolved
with the atmospheric smearing function.

The distribution of $\mu_0$ values indicates the presence of two subsamples:
those with $\mu_0$ values greater than 23.5 $mag/arcsec^2$ which we defined as
LSB and those with higher surface brightness designated HSB in this work.
Note that all galaxies of the faintest surface brightness class (according
to the RFGC) were included to our LSB objects.

Although our sample enables us to compare of the photometric parameters of
LSB and HSB disks, the sample is very limited. We also incorporated one more
sample of edge-on galaxies whose photometric parameters have been published
by Barteldrees etc. [4]. The authors used the similar passband and the same
fitting functions to extract the photometric parameters.

\section{LSB versus HSB: the vertical scale height of galactic disk\\ as a new
feature to compare.}

As it was shown in [2] and [5], the galaxies of lower surface brightness
tend to show the higher $z_0/h$ ratio. However, mostly HSB galaxies were
considered in those previous studies. Here we show the dependence of $z_0/h$
on the central surface brightness $\mu_0$ for our sample. Our galaxies are
denoted by the squares in Fig.1. 
The open squares are for the HSB subsample
whereas the filled ones designate the LSB galaxies. The galaxies taken from
[4] are plotted as crosses. Futhermore, the near-infrared (K band) sample of
edge-on galaxies from [5] is available for the comparison (the 2MASS sample
hereafter). The systematic difference between the R and K images in $z_0/h$
(1.4 times, see [6] for details), in the surface brightness (2.4 $mag$, see
[7]), and the internal extinction (1.2 $mag$) were taken into account.

Fig.1 shows all three samples together. The 2MASS sample is denoted by the
small filled triangles. A trend in Fig.1 is seen well, 
in average of 2 mag
difference in $\mu_0$ leads to 1.5 change in the ratio of the scales. Note that
there is no correlation seen when $h$ or $z_0$ were drawn against $\mu_0$
separately.

Following [6], we calculated the ratio of total mass $M_t$ to B-band disk
luminosity $L_B$: $M_t~=~0.5G ~4h~{V_m}^2/L_B$. Here $V_m$ is the maximum of
rotation curve (taken from the LEDA). The value of $L_B$ is inferred from the
absolute B-magnitude of a galaxy taken from the LEDA and corrected for the
internal absorption. In Fig. 3a one can see the values of $M_t/L_B$ plotted
versus the ratio $z_0/h$. The notation in the figure is the same as in
Fig.1.
The three curves in Fig.3a are the same as in Fig.2 which were recalculated
for the values of the mass to luminosity ratio $M/L$ of 1, 2 and 4. As it is
seen in Fig.3a, most of the galaxies have a reasonable value of $M/L$ between
1 and 4. The mass to luminosity ratio is higher systematically for our LSB
galaxies in comparison with HSB ones.

Since the luminosity ratio $Lb/Ld$ reflects the mass ratio, the mass of the
dark halo $M_h$ can be estimated from the dependence in Fig. 3. We drew the
ratio of dark to luminous matter $M_h/(M_b + M_d)$ for our galaxies ishown n
Fig.3b. Here we designate the bulge mass as $M_b$. We show the ratio of "dark"
to "luminous" mass $M_h/(M_b + M_d)$ for our galaxies. Here we designate the
bulde mass as $M_b$. We assume that the bulge and the disk have roughly the
same $M/L$. Surprisingly, there is no systematic difference between the values
of $M_h/(M_b + M_d)$ for the galaxies of different central surface brightness.

We also enable to compare the mass of the spherical subsystem $M_s$ for our
galaxies. In Fig. 3c we present how the ratio $M_s/M_d$ depends on the disk
central surface brightness. We kept the same notation as in Fig. 1 and Fig.
3a. The figure indicates that the LSB galaxies do not have the most massive
dark matter halos, but have the most massive spherical subsystem. That
supports the result by Graham ([10]) that not all LSB galaxies are the
dark-matter dominated objects.

\section{Conclusions}

\noindent 1) We present the results of the photometric observations made for the
sample of edge-on galaxies. Our sample contains four LSB galaxies as well as
seven HSB ones. The photometric disk scales (vertical and radial), the disk
central surface brightness and the bulge to disk luminosity ratio were
determined.

\noindent 2) Stellar disks of our LSB galaxies are thinner than HSB ones. There is a
good correlation between their central surface brightness and their vertical
to radial scales ratio.

\noindent 3) Our LSB galaxies tend to harbor the massive spherical subsystems as well
as to havehigher values of the mass-to-luminosity ratio in their disks when
compared to the HSB objects. Nevertheless, the dark halo is not strictly the
most massive subsystem in our LSB galaxies. The LSB galaxies appear to be
the spherical dominated systems, but not the "dark matter dominated" ones.

\begin{acknowledgments}
\noindent Acknowledgments

\noindent D.B. is supported by NASA/JPL through grant NRA-99-04-OSS-058. The project
was partially supported by Russian Foundation for Basic Research via grant
01-02-17597. We have made use of the LEDA database.
\end{acknowledgments}

\bigskip
REFERENCES
\bigskip

{\small
\noindent [1]  Karachentsev, I., Karachentseva, V., Kudrya, Y., et al.//       
     1999, Bull. Of Special Astrophys. Obs., 1999, 47, 5, (RFGC)

\noindent [2]  Bizyaev, D.//  2000, Astron. Lett. 26, 219.

\noindent [3]  van der Kruit, P., Searle, L. // 1981, A\&A, 95, 105 and 116

\noindent [4]  Barteldrees, A., Dettmar, R.-J. // 1994, A\&AS, 103, 475

\noindent [5]  Bizyaev, D., Mitronova, S. // 2002, A\&A, 389, 795 

\noindent [6]  Zasov, A., Bizyaev, D., Makarov, D. etc. // 2002, 
Astron.Lett. 28, 527

\noindent [7]  de Jong, R. // 1996, A\&A, 313, 45

\noindent [8]  Zasov, A., Makarov, D., Mikhailova, E. // 1991, Astron.Lett., 
17, 374

\noindent [9]  Mikhailova, E., Khoperskov, A., Sharpak, S. // 2001, in 
Stellar Dynamics: From Classic to Modern", Ed. by L.Ossipkov     
and I.Nikiforov, p. 147

\noindent [10] Graham, A. // 2002, MNRAS, 334, 721
}

\clearpage

\section*{FIGURE CAPTIONS}

\noindent {\bf Fig.1.} The vertical to radial scale ratio $z_0/h$ is shown against the disk
central surface density $\mu_0$. The open squares are for the HSB subsample,
the filled ones designate our LSB galaxies. The galaxies taken from [4] are
shown by crosses, the 2MASS sample is designated by the small filled
triangles.

\noindent {\bf Fig.2.} The ratio od scales $z_0/h$ is plotted versus the relative mass
of the spherical galactic subsystem $M_s/M_d$ according to the numerical
simulations described in [9].

\noindent {\bf Fig.3.} a) The scales ratio $z_0/h$ and the total mass to disk
luminosity rati $M_t/L_B$. The same notation applies here as in Fig.1. The
three curves show the model results (see Fig.2) recalculated for $M/L$
values of 1 (solid curve), 2 (dash-dotted) and 4 (dashed). The value of M/L
is systematically higher for our LSB galaxies.

\noindent b) The ratio of dark to luminous mass $M_h/(M_b + Md)$ is plotted against
the disk central surface brightness $\mu_0$ with the same notation as in
Fig. 3a. There is no systematic difference in the relative dark matter mass
between the LSB and HSB galaxies.

\noindent c) The ratio of the spherical to disk mass $M_s/M_d$ is shown in dependence
on the disk central surface brightness. The LSB galaxies have the more
massive spherical subsystem when compared to HSB ones.

\clearpage

\begin{figure}
\plotone{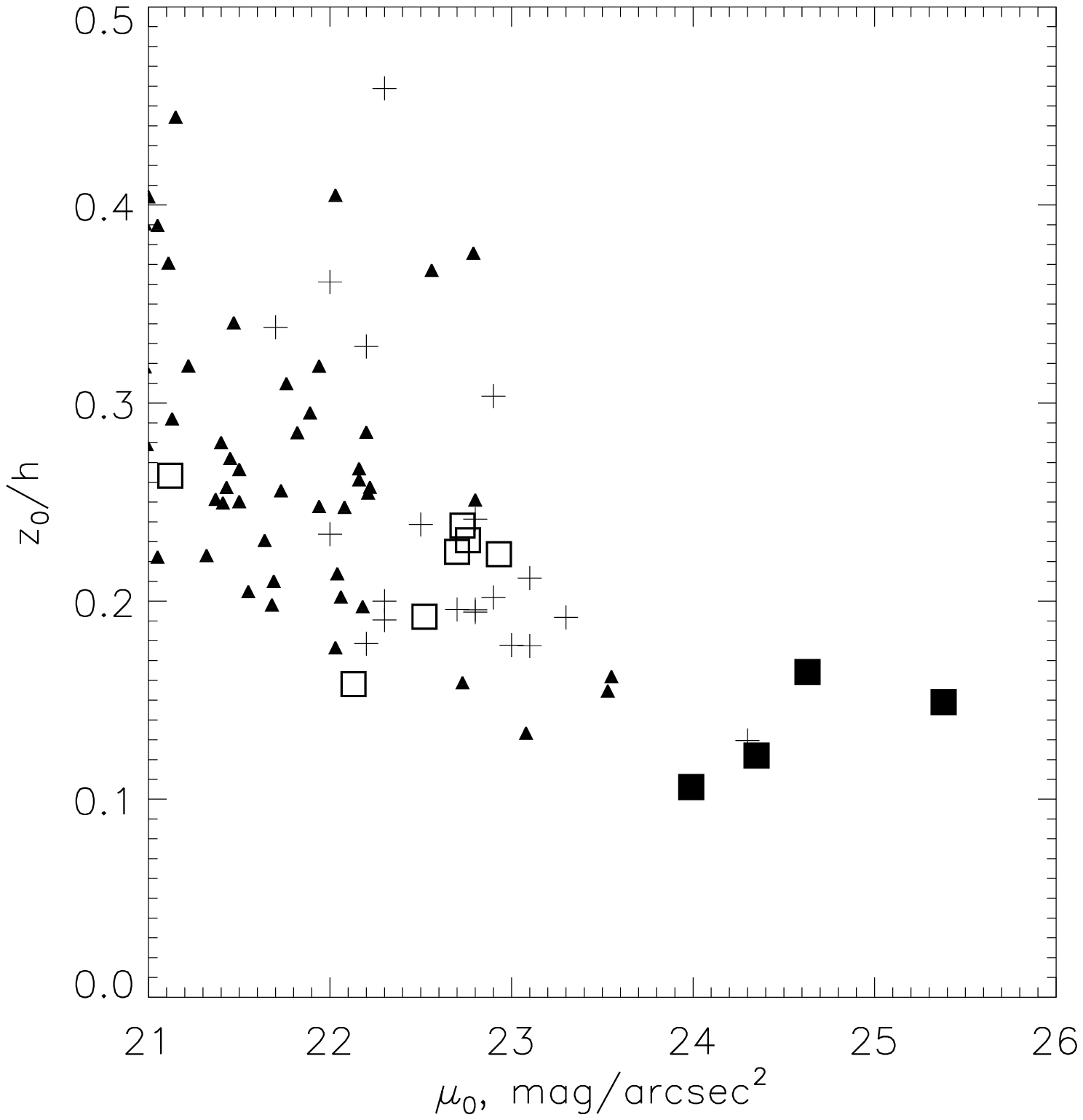}
\caption{.
\label{fig1}}
\end{figure}

\clearpage

\begin{figure}
\plotone{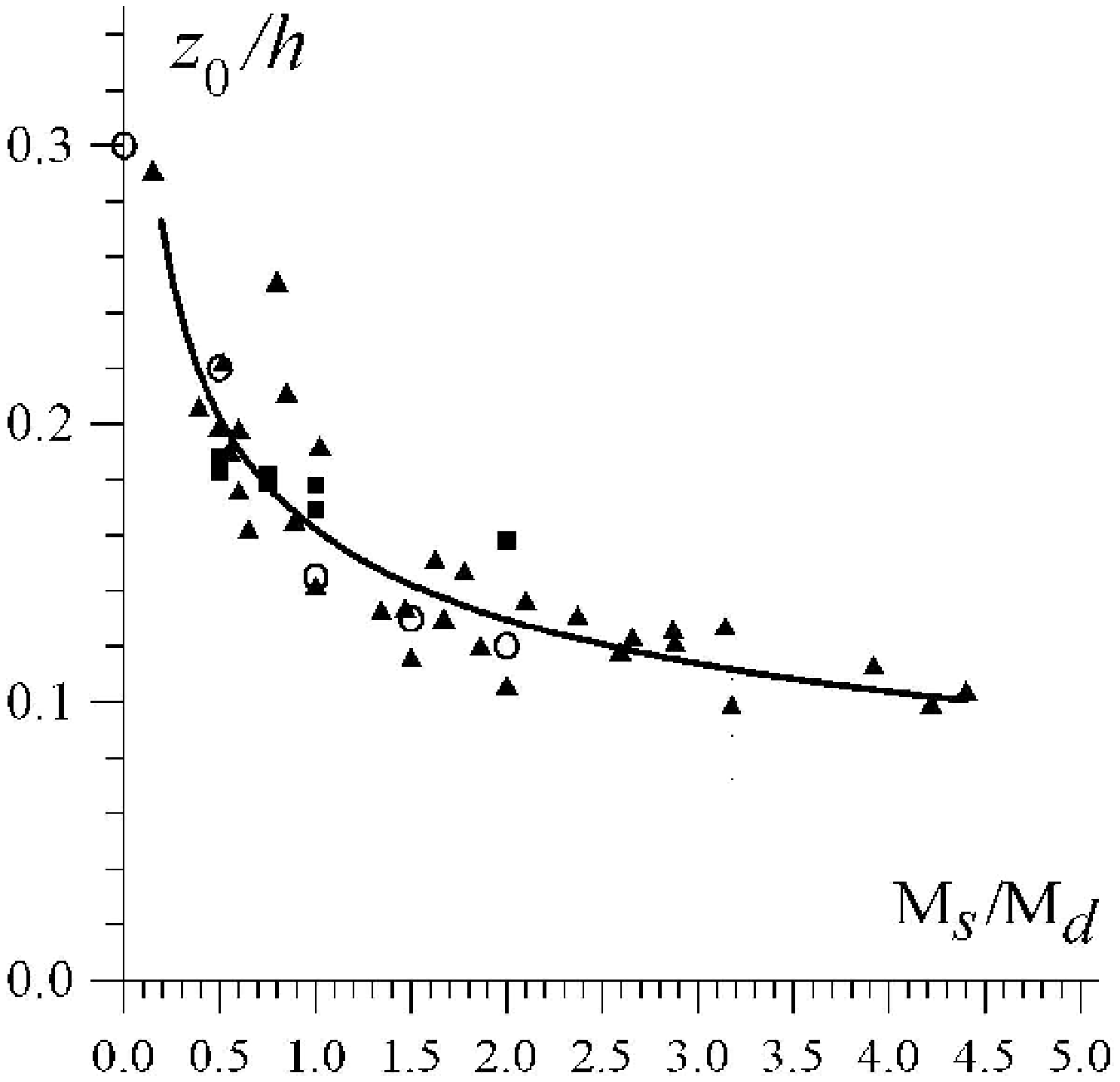}
\caption{.
\label{fig2}}
\end{figure}

\clearpage

\begin{figure}
\plotone{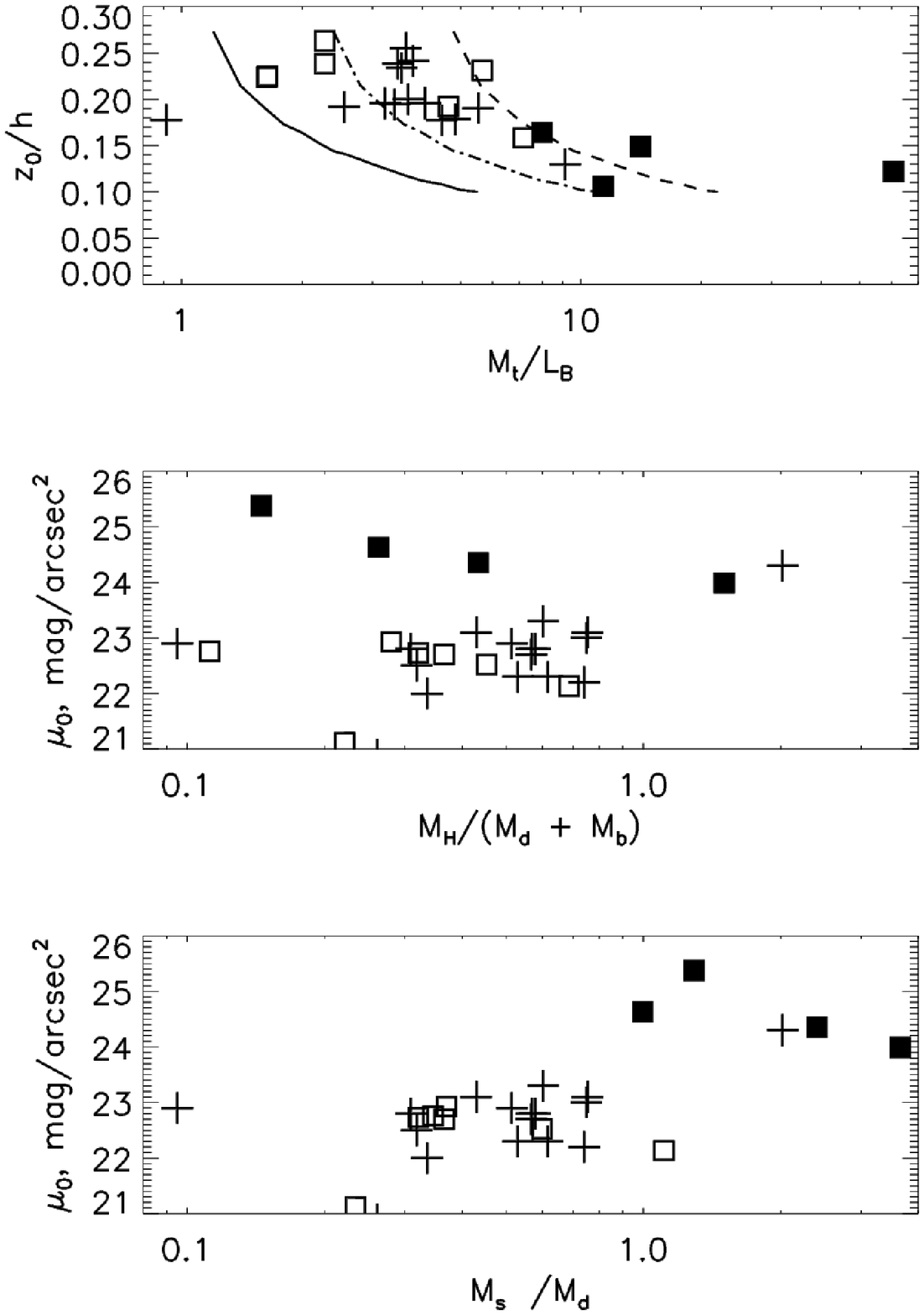}
\caption{.
\label{fig2}}
\end{figure}

\end{document}